\documentclass[
    ,final            % use final for the camera ready runs
  ]
  {aipproc}

\layoutstyle{8x11double}

\begin{document}

\title{VLT observations of Compact Central Objects}


\author{R.P. Mignani}{
  address={University College London - Mullard Space Science Laboratory, Holmbury St. Mary, Dorking, RH56NT, UK}
}

\author{S. Zaggia}{
  address={INAF - Osservatorio Astronomico di Padova, Vicolo dell'Osservatorio 5, Padua, I-35122, Italy}
}

\author{D. Dobrzycka}{
  address={ESO - European Southern Observatory,
Karl Schwarzschild Str. 2, Garching, D-85740, Germany}
}    

\author{G. Beccari}{
  address={INAF - Osservatorio Astronomico di Bologna, Via Ranzani 1, Bologna, I-40127, Italy}
}

\author{A. de Luca}{
  address={INAF - Istituto di Astrofisica Spaziale e Fisica Cosmica di Milano,
Via Bassini 15, I-20133 Milano, Italy}}

\author{S. Mereghetti}{
  address={INAF - Istituto di Astrofisica Spaziale e Fisica Cosmica di Milano,
Via Bassini 15, I-20133 Milano, Italy}}

\author{P. A. Caraveo}{
  address={INAF - Istituto di Astrofisica Spaziale e Fisica Cosmica di Milano,
Via Bassini 15, I-20133 Milano, Italy}}

\begin{abstract}
We present  the first results of  our VLT observation  campaign of the
Central  Compact  Objects (CCOs) in SNRs RX  J085201.4-461753  (Vela Jr),  1E
1648-5051 (RCW  103) and  RX J171328.4-394955 (G347.3-0.5).   For Vela
Jr., we found that the  source is embedded in  a compact optical
nebulosity, possibly a bow-shock  or a photo-ionization nebula, and we
 identified a candidate IR counterpart to the CCO. For RCW 103, we
found no convincing  evidence neither for 6 hrs  IR modulation nor for
variability on any  time scale from the proposed  counterpart, as well
as for the other candidates close to the revised Chandra position. For
G347.3-0.5, we  identified few possible IR  counterparts but none
of them is apparently associated with the CCO.
\end{abstract}

%%%%%%%%%%%%%%%%%%%%%%%%%%%%%%%%%%%%%%%%%%%%%%%%%%%%%%%%%%%%%%%%%%%
%%
%% The below \maketitle command inserts the actual front matter data.
%% It has to follow the above declarations.
%%
%%%%%%%%%%%%%%%%%%%%%%%%%%%

\maketitle

%%%%%%%%%%%%%%%%%%%%%%%%%%%%%%%%%%%%%%%%%%%%
%% MAINMATTER
%%
%%%%%%%%%%%%%%%%%%%%%%%%%%%%%%%%%%%%%%%%%%%%%%%%%%%%%%%%%%%%%%%%%%%%%%%%%%%%
%% Headings:
%%
%% The aipproc supports three heading levels, i.e., \section,
%%	\subsection, and \subsubsection.
%%%%%%%%%%%%%%%%%%%%%%%%%%%%%%%%%%%%%%%%%%%%%%%%%%%%%%%%%%%%%%%%%%%%%%%%%%%%
%% Cross-references:
%%
%% Page numbers (\pageref) and headings can NOT be referenced in the class,
%% since before being produced, no page numbers are determined.
%%
%% Tables, figures, and equeations can be referenced by using the LaTex
%% 	commands \label and \ref. For references to equation numbers, \eqref
%%	can be used, which will print "(1)" (while \ref will result in "1").
%%
%%%%%%%%%%%%%%%%%%%%%%%%%%%%%%%%%%%%%%%%%%%%%%%%%%%%%%%%%%%%%%%%%%%%%%%%%%%%
%% Lists: 
%%
%% Standard "itemize", "enumerate", etc. list environments are supported.
%%%%%%%%%%%%%%%%%%%%%%%%%%%%%%%%%%%%%%%%%%%%%%%%%%%%%%%%%%%%%%%%%%%%%%%%%%%%
%% Urls:
%%
%% \url{} command is provided for documenting URLs.
%%%%%%%%%%%%%%%%%%%%%%%%%%%%%%%%%%%%%%%%%%%%

\section{Introduction}

X-ray observations have unveiled the existence of enigmatic point-like
sources at  the centre of  young SNRs.  These sources,  dubbed Central
Compact Objects  (CCOs), are thought  to be neutron stars  produced by
the  supernova explosions.  However,  their X-ray  phenomenology, with
scanty evidence  for pulsations, thermal X-ray radiation,  and lack of
magnetospheric  activity,  make   them  markedly  different  from  the
majority  of  young  neutron  stars.   Their radio  silence  is  still
unexplained, though.   The recent discovery of a  few hours (orbital?)
periodicity and  long term variability,  make these objects  even more
intriguing.  As  a part  of a coherent  multi-wavelength observational
program,  we   have  performed  VLT  observations   of  the  candidate
counterpart to the RCW103 CCO, and  for the Vela Jr and the G347.3-0.5
CCOs, for which no deep optical/IR data were available.

\section{Vela Jr.}

R-band  observations of  the Vela  Jr.  CCO have  been performed  with
FORS1(0".1/px).   Data  have  been  reduced/calibrated with  the  FORS
pipeline  and  the astrometry  computed  with  2MASS  (0.12'' rms).  A
4"$\times$4" unresolved nebula (Mignani et al. 2007a,b) is detected at
the  CCO X-ray  position (Fig.  1) but  no point-like  source  down to
R$\sim$5.6  (3 $\sigma$).  Both the  nebula's flux  and  structure are
similar to the  H$\alpha$ ones (Fig. 1), suggesting  that its spectrum
is  dominated by  pure H$\alpha$  line emission.  This means  that the
nebula is most likely a  bow-shock produced by the neutron star motion
through the  ISM or, alternatively, a  photo-ionization nebula powered
by  UV radiation  from  a  hot neutron  star.   A synchrotron  nebula,
powered by  the relativistic particle  wind from the neutron  star, is
the  less likely interpretation  because of  its non-detection  in the
X-rays  and of  the apparent  lack of  continuum emission  (Mignani et
al. 2007b). High-resolution imaging observations, now in progress with
the  HST, will  help to  better assess  the nebula  morphology  and to
identify its nature.

IR (JHK$_s$) observations have  been performed with NACO (0.027''/px).
Data  have been  reduced/calibrated  with the  NACO  pipeline and  the
astrometry computed with 2MASS (0.13''  rms). A faint object (Fig.  2)
is identified near  the Chandra error circle (Mignani  et al.  2007b).
Daophot photometry yields J $>$22.6, H$\sim$21.6, K$_s$$\sim$21.4.  No
evidence for  short-term variability has been  found.  The candidate's
location  in  the  J,H-K$_s$  color  Magnitude Diagram  (CMD)  is  not
significantly different from those of field stars (Fig. 3).  If object
A  were the  CCO neutron  star, its  IR flux  would be  $\sim$10 times
higher  than expected  for a  1 kpc  distance and  for a  Vela-like IR
emission efficiency. If the emission  from object A were ascribed to a
fallback disk around the CCO  neutron star, its IR-to-X-ray flux ratio
($\sim 5  \times 10^{-4}$) would be  similar to the  value measured for
the AXP 4U0142+61. Finally, if object A were a star in a binary system
with the CCO neutron star, its colors would be consistent with a mid M
type at d$\sim$2 kpc, or with an L dwarf.

\begin{figure}
\centering
  \includegraphics[bb=70 150 570 650,height=6cm,angle=0,clip]{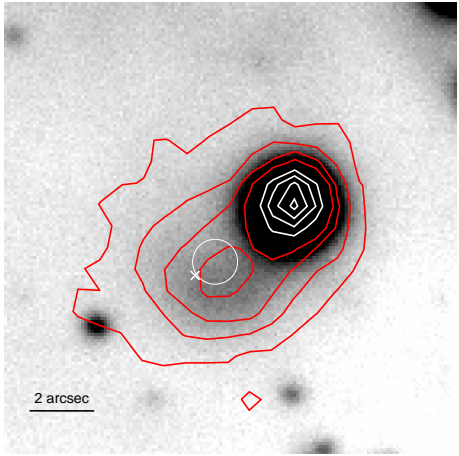}
\caption{VLT/FORS1 image (R-band) of the Vela Jr. CCO field. The  circle (0.7'') shows the Chandra 
X-ray position while the cross marks the position of object ``A'' (see Fig.2). The isophotal contours from the UKST H$\alpha$ image are overlaied.}
\end{figure}

\begin{figure}
\centering
  \includegraphics[bb=50 140 250 340,height=6cm,angle=-90,clip]{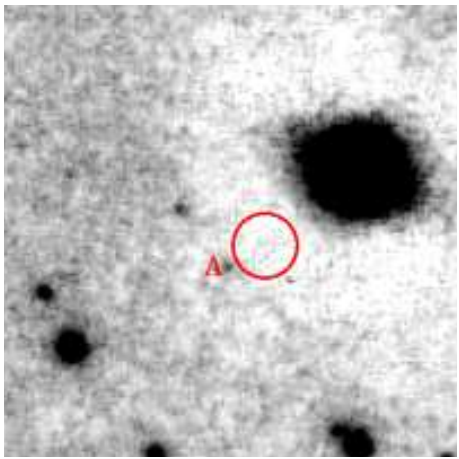}
  \caption{VLT/NACO image (K$_s$-band) of the Vela Jr.CCO field. The Chandra 
X-ray position is marked by the circle (0.7''). The CCO candidate IR counterpart is labeled 
(``A'').}
\end{figure}

\begin{figure}
  \includegraphics[bb=30 160 570 700,height=6cm,angle=0,clip]{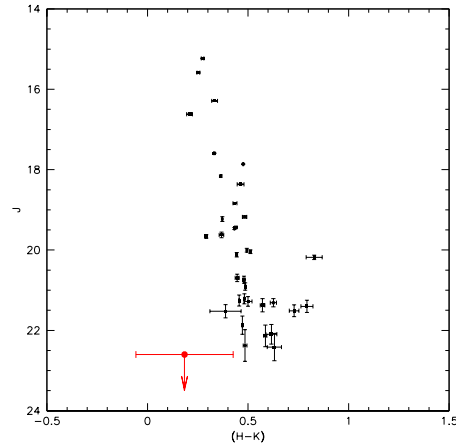}
  \caption{J,H-K$_s$ CMD of all the stars identified in the VLT/NACO field. Object A  marked in red.                                        }
\end{figure}

\section{RCW103}

K$_s$-band observations of the RCW103 CCO have been obtained with NACO
in 2006 on two consecutive nights to monitor the candidate counterpart
(Sanwal et  al.  2002) along  the 6 hrs  X-ray cycle. Fig.4  shows the
updated Chandra CCO position (De  Luca et al.  2007) registered on the
NACO   co-added   image.    The   proposed  counterpart   (object   1;
K$_s$$\sim$18) falls off the X-ray  position. No clear evidence of a 6
hrs  modulation  is found.   This  is  confirmed  by the  analysis  of
archival 2001 VLT/ISAAC H-band  and 2002 HST/NICMOS 160W observations.
From the VLT  and HST datasets we found no  evidence of variability on
months or years time scales either.  Furthermore, the colors of object
1 (Fig. 5) are not peculiar wrt those of the field stellar population.
Other fainter  ($18<K_s<20.4$) candidates  (3-7) are visible  close to
the Chandra position (Fig. 6). However, also for them no evidence of a
6 hrs modulation is found neither in the 2006 VLT/NACO nor in the 2002
HST/NICMOS 160W  observations.  No other candidate is  detected in the
co-added  NACO image close  to the  X-ray error  circle down  to K$_s$
$\sim$22.1.   We  have used  the  archival  HST/NICMOS  160W and  205W
observations to search for possible long term variability and to study
the colors of the new candidates  (all too faint to be detected in the
VLT/ISAAC  observations).  However,  no evidence  for  short/long term
variability has been found for any of the new candidates. In addition,
none of them shows evidence for peculiar colours (Fig.7) wrt the field
stellar population.  Thus, based on  both colours and on  the apparent
lack of  variability on any time  scale, we conclude that  none of the
present candidates can be spotted as the likely CCO counterpart, which
remains still unidentified.

\begin{figure}
\centering
  \includegraphics[bb=15 125 245 375,height=6cm,angle=-90,clip]{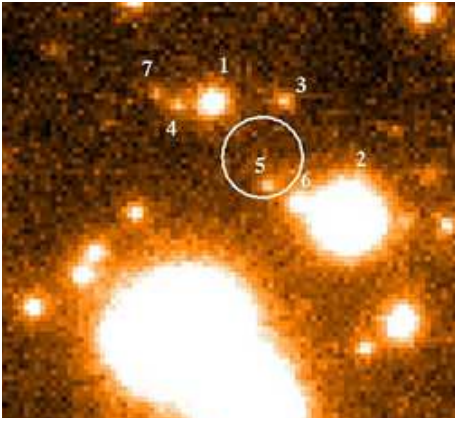}
\caption{VLT/NACO  image (K$_s$-band) of the RCW 103 CCO field. The circle (0.4'') corresponds to 
the updated Chandra CCO position. Object 1 is the candidate CCO 
counterpart of Sanwal et al. (2002). }
\end{figure}

\begin{figure}
\centering
  \includegraphics[bb=70 160 570 710,height=7cm,angle=0,clip]{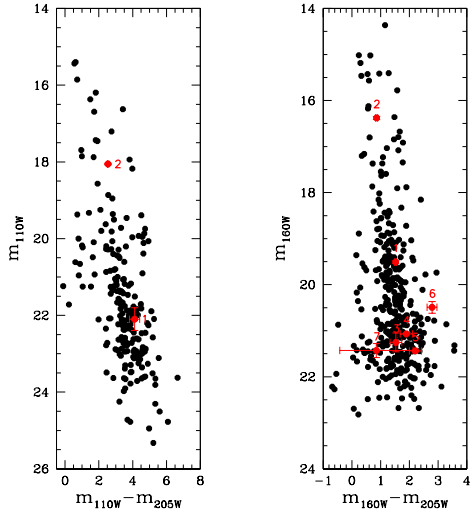}
  \caption{110W,110W-205W and 160W,160W-205W CMDs of all the stars detected in the HST/NICMOS 
field. All candidates of Fig. 4 are marked in red and labeled.   
}
\end{figure}

\section{G347.3-0.5}

H (2400s)  and K$_s$ (4600s)  band observations of the  G347.3-0.5 CCO
have been  performed with NACO  on two consecutive nights  (Mignani et
al. 2007c). Data  reduction and analysis has been  performed as above.
Five, possibly  six objects ($19.3<H<21.1$), are  identified within or
close the Chandra  X-ray position (Fig. 7, left).   None of them shows
evidence  for  significant  variability  on  short  time  scales.   In
addition,  none  of  them  is  characterized by  extreme  or  peculiar
colors. Indeed, their  locations in the H, H-K$_s$  CMD (Fig.7, right)
fall  right on  the sequence  of  the field  stars. The  colors of  the
candidates  are very  red (H-K$_s$$\sim$1.5)  and are  consistent with
those of stars more absorbed  ($N_H > 5 \times 10^{21}$ cm$^{-2}$), and
likely more  distant (d$>$6kpc), than  the CCO.  A very  late spectral
type star  (e.g.  an  L dwarf) could  have an H-K$_s$  compatible with
those of  the candidates but it should  be no further than  0.5 kpc to
have the  same brightness.   At variance with  the Vela Jr.   case, we
thus regard  it as  unlikely that  the CCO has  a companion  star.  No
other candidate  is detected close to  the X-ray error  circle down to
H$\sim$22 and  K$\sim$20.5. We note  that, given the CCO  distance and
the  intervening  interstellar  extinction,  these  upper  limits  are
largely consistent with emission from an isolated neutron star or from
a fallback disk.

\begin{figure}
  \includegraphics[bb=35 85 285 335,height=6cm,angle=-90,clip]{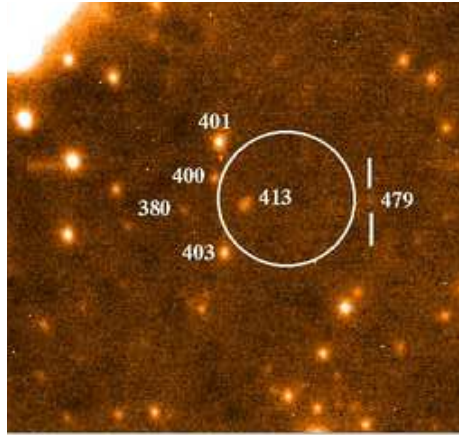}
  \caption{VLT/NACO image (H-band) of the G347.3-0.5 CCO. The circle (1.0'') marks
 the Chandra  position. Possible counterparts are labeled, with object 479 detected only 
at $<5 \sigma$. }
\end{figure}

\begin{figure}
  \includegraphics[bb=30 160 570 700,height=6cm,angle=0,clip]{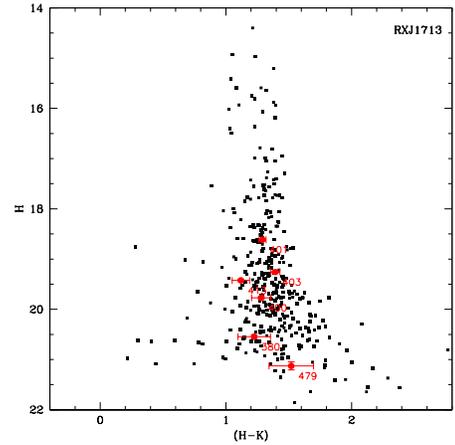}
  \caption{H,H-K$_s$ CMD of all stars detected in the VLT/NACO field. All candidates of Fig. 5 are marked in red and labeled.}
\end{figure}

\begin{theacknowledgments}
 RPM thanks N. Ageorges (ESO) and the UCL Graduate School and the Royal Society for financial support.
\end{theacknowledgments}

%%%%%%%%%%%%%%%%%%%%%%%%%%%%%%%%%%%%%%%%%%%%%%%%
%% The bibliography can be prepared using the BibTeX program or
%% manually.
%%
%% The code below assumes that BibTeX is used. Compliant BibTex styles
%% are aipproc (for use with natbib) and aipprocl (if natbib is missing
%% at the site).
%%
%% Please run "bibtex \jobname" to obtain the bibliography and 
%% then re-run LaTeX twice to fix the references!
%%
%% When referring to citations in the text, in quare brackets [] show
%% the number in order of appearance. References in the References
%% section are listed in the same numerical order.
%%%%%%%%%%%%%%%%%%%%%%%%%%%%%%%%%%%%%%%%%%%%%%%%%

\bibliographystyle{aipproc}   % if natbib is available
%\bibliographystyle{aipprocl} % if natbib is missing

%%%%%%%%%%%%%%%%%%%%%%%%%%%%%%%%%%%%%%%%%%%
%% You probably want to use your own bibtex database here
%%%%%%%%%%%%%%%%%%%%%%%%%%%%%%%%%%%%%%%%%%%

%\bibliography{sample}

\begin{thebibliography}{9}

\bibitem{adl} De Luca, A., Mignani, R.P., Zaggia, S., et al., 2007, ApJ, submitted

\bibitem{mig07a} Mignani, R.P., Bagnulo, S., De Luca, A., et al. 2007, Ap\&SS, 308, 203

\bibitem{mig07b} Mignani, R.P., De Luca, A., Zaggia, S., et al., 2007a, A\&A, in press

\bibitem{mig07c} Mignani, R.P., Zaggia, S.,  De Luca, A., 2007c, A\&A, submitted

\end{thebibliography}

%%%%%%%%%%%%%%%%%%%%%%%%%%%%%%%%%%%%%%%%%%%%%%%%%
%% If the bibliography is
%% produced without BibTeX, comment out the above lines, use
%% \begin{thebibliography}{widest-label} environment to hold 
%% the list of references and 
%% \bibitem{label} command to start a bibliographical entry having
%% the "label" for use in \cite commands.
%%
%% For your convenience a manually coded example is appended
%% after the \end{document}
%%%%%%%%%%%%%%%%%%%%%%%%%%%%%%%%%%%%%%%%%%%%%%%%

%%%%%%%%%%%%%%%%%%%%%%%%%%%%%%%%%%%%%%%%%%%
%% The following lines show an example how to produce a bibliography
%% without the help of the BibTeX program. This could be used instead
%% of the above.
%%%%%%%%%%%%%%%%%%%%%%%%%%%%%%%%%%%%%%%%%%%

\end{document}